# A cost-effective quantum eraser demonstration

Aarushi Khandelwal[1], Jit Bin Joseph Tan[3], Tze Kwang Leong[4], Yarong Yang[4], T Venkatesan[1,2,†], and Hariom Jani[1,*]

[1]NUS Nanoscience and Nanotechnology Initiative, National University of Singapore, Singapore
[2]ECE Department, National University of Singapore, Singapore
[3]Physics and Engineering Department, NUS High School of Math and Science, Singapore
[4]Sciences Branch, Curriculum Planning and Development Division, Ministry of Education, Singapore
[†] venky@nus.edu.sg
[*] hariom.k.jani@u.nus.edu

**Abstract**

The quantum eraser is a variation of the celebrated Young's interference experiment that can be used to demonstrate the elusive complementarity principle in quantum physics. Here we show the construction of its classical analogue for deployment in classrooms in a simple, cost-effective yet robust manner by employing a laser pointer, double-slits, and polarizers.

## 1. Introduction

The quantum eraser is an interferometric experiment that demonstrates the dual nature of light with photons passing through a double-slit setup exhibiting either wave-like (i.e. interference) or particle-like (i.e. 'which- way' information) behaviour, depending on whether the setup allows optical paths to be indistinguishable or distinguishable, respectively. The choice lies in the hands of the experimenter. Interestingly, one may add optical components downstream to allow the experimenter to 'erase' the previous choice of path (in)distinguishability, reversing the observed behaviour of light[1,2]. This experiment can serve as a simple yet powerful demonstration of the otherwise abstract concept of complementarity for students in high schools or at an introductory undergraduate level.

Ideally, the quantum eraser experiment should be performed with a single-photon light source. However, due to the need for specialized and relatively expensive equipment as well as long experiment durations required to obtain sufficient statistics, it is not appropriate for a classroom demonstration[3,4,5]. Instead, replacing the single-photon source with a laser and controlling path (in)distinguishability via polarizers can produce similar effects, almost instantly, making them more suitable for deployment in classrooms[6]. Such a setup would serve as a classical analogue of the quantum eraser.

The setup in Ref. 6 creatively uses a blocking wire clasped between two polarizers to create this experiment. However, due to lack of control over dimensions, we found it difficult to make this setup reproducibly and robustly. In addition, the interference pattern can undergo distortion due to the light deviation from polarizer edges adjacent to the blocking wire and students are not usually familiar with the diffraction patterns obtained from wire diffraction.

Hence, in this article, we present a new method of building a more controlled and robust version of the classical analogue of the quantum eraser that employs in-syllabus concepts (at the A-level or equivalent standards) such as double- slit and single-slit diffraction, and Malus law, which we prepared for implementation in Junior Colleges in Singapore[7]. This demonstration was scaffolded by a predict–explain–observe–explain styled pedagogy to enhance student engagement and learning, see supplementary information for details.

## 2. Experimental Setup

Three components are required to construct this demonstration (schematic in Figure 1). (i) A coherent light source, i.e. a red laser. (ii) Set of polarized double-slits laser-cut into a cardboard, where the polarizers are placed carefully to ensure that the optical path is neither blocked nor distorted (Figure 2(a)). Two configurations of polarized double-slits are required, i.e. VH and VV polarized slits, which refer to one vertical (0°) and one horizontal (90°) and two vertical polarizers placed on the slits, respectively. (iii) An analyser cup made of a polarizing strip attached at the bottom of a styrofoam cup. The V and H directions are carefully labelled corresponding to the polarizers on the slits (Figure 2(b)). The overall assembly does not necessarily require an optical bench and can be set directly on a table (Figure 2(c)).

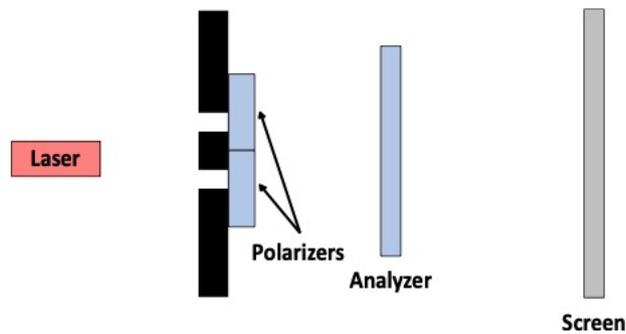

*Figure 1: Schematic of the overall quantum eraser set-up with a laser, polarized double-slits, an analyser, and a screen placed ∼2–3 m away.*

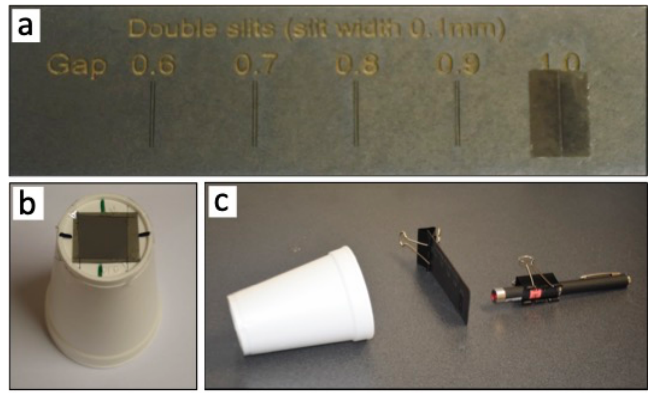

*Figure 2: The demonstration (a) the double-slit setup prepared by laser-cutting a solid piece of cardboard to allow accurate control over the width of the slits (∼0.1 mm) and separation between the centres of the slits (∼0.6–1 mm). This is necessary to observe clear interference patterns on the screen. Polarizing filters are carefully taped in front of the slits. (b) The analyser made using a styrofoam cup and polarizing film attached to a square cut-out (∼2.5 × 2.5 cm$^2$). (c) The overall set-up. Regular fasteners are used to hold the components upright and the experiment should be performed in a dark room to increase the visibility of the patterns.*

In each experiment, conducted separately for VV and VH configurations, the laser is switched on and then the analyser cup is rotated to study its effect on the pattern on the screen. It should be noted that the laser light incident on the polarized double-slit setup should be oriented diagonally (i.e. 45° or 135° polarization) to allow equal transmission of light through both V and H slits. This is achieved by either rotating the laser or placing an additional diagonal polarizer between components (i) and (ii).

## 3. Results
### *3.1 VV polarized slits*

Photons passing through the slits in this configuration would undergo the same degree of polarization whether they go through the left- or right-side slit, rendering the two paths indistinguishable. Hence, one would always observe a double-slit interference pattern on the screen as 'which-way' information cannot be determined. Rotating the analyser cup only changes the total intensity of the pattern, which is maximum for the 0° orientation (as both slits are V polarized) and a minimum for the 90° orientation, following Malus law (Figure 3(a)). Using HH polarized slits instead would yield the same results, with the exception of a 90° angular phase shift.

### *3.2 VH polarized slits*

The photons' paths through the two slits in this configuration are now physically distinguishable due to the different polarizations. Hence, placing the analyser cup in the 0° or 90° orientations would selectively only allow photons going from the left- or right-side slit, respectively, thereby yielding 'which-way' information. Instead of the double-slit interference pattern, one would now observe a single-slit interference pattern from photons passing through the slit whose polarization matches the analyser (Figure 3(b)). However, when the analyser cup is oriented diagonally (45° or 135°) the distinguishability of photons' paths through the two slits is erased, making it impossible to determine which slit the photons passed through. Hence, the double-slit interference pattern reappears (Figure 3(b)). Placing the analyser cup at an angle away from the above orientations would result in a reduced visibility of interference fringes.

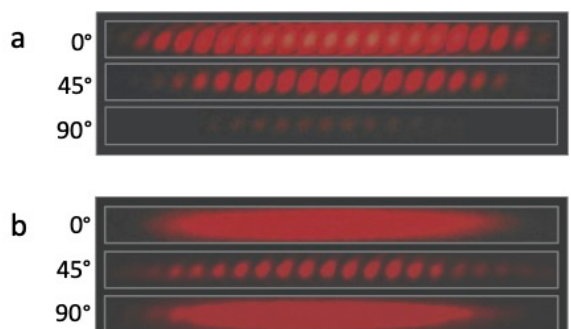

*Figure 3: The intensity patterns at 0°, 45°, and 90° orientation of the analyser cup, with the polarized slits in the (a) VV and (b) VH configurations.*

## 4. Classroom discussion—classical and quantum pictures

The experimental patterns resulting from a laser source, discussed above, can be predicted by a purely classical wave picture[6]. In the VH configuration, an analyser aligned with either one of the slits (0° or 90°) blocks all light from the other one, producing only a single-slit pattern. However, a 45° analyser allows light from both V and H slits to pass and interfere, reproducing the double-slit interference pattern.

To help the students appreciate the quantum interpretation of the observed results (involving wave-particle duality and the quantum eraser effect), the educator may follow up the demonstration with a classroom discussion on what would happen if the laser source were replaced by a single photon source. Such a discussion would explore how the probability distributions of the single photon trajectories (as per the Born interpretation) integrated over a long time would produce results similar to those in Figure 3 and section 3 [1,2,6,8,9]. Overall, this discussion would help students to better understand the nature of science, where multiple models are used to understand the same phenomenon (in this case, the nature of light).

## 5. Conclusion

A classical analogue of the quantum eraser experiment presented in this paper is a cost-effective yet robust demonstration, enabling high school and introductory undergraduate students to gain a hands-on experience of the complementarity principle.

## Acknowledgments

This MAF study was funded by the Education Research Funding Programme, National Institute of Education (NIE), Nanyang Technological University, Singapore, project number AFD 04/16 VV and the experiments were scaled up to all Junior Colleges in Singapore with the support of AST. We thank Dr D Wong and Dr D Tan for their support on this project. We also thank Dr A Patra and Dr A Srinivasan for discussion. The views expressed in this paper are the authors' and do not necessarily represent the views of the host institution.

# Supplementary information

## 1. Introduction for intellectual scaffolding

When delivering the demonstration in classrooms, sufficient intellectual scaffolding prior to the experiment will enable students to better grasp the key learning objectives of the demonstration[R1,R2]. We anticipate that some familiarity with the following concepts and ideas would be beneficial:

- The superposition principle and the wave nature of light.
- The effect of polarizers on light and Malus Law.
- The behavior of classical waves passing through 2 slits exhibiting interference[R3].
- The behavior of classical particles passing through 2 slits exhibiting which-way information[R3].
- The complementarity principle.
- The idea that a photon passing through a double-slit can exhibit particle-like or wave-like behaviors depending on whether or not one can obtain which-way information.

## 2. PEOE Questionnaires

In a classroom demonstration delivery, a series of 5 questionnaires can be given to the students to stimulate their understanding of the experiment in a Predict-Explain-Observe-Explain (PEOE) framework, which are included here as an example:

- Pre-demo questions: *To help clear conceptual issues before the experiment*
- Experiment Part 1 (VV slits).
- Experiment Part 2 (VH slits and vertical analyzer).
- Experiment Part 3 (VH slits and diagonal analyzer).
- Post-demo questions: *To help evaluate the level of understanding after the experiment*

Our model answers to the MCQ questions are highlighted below.

### Questionnaire 1: Pre-demo questions

1. When vertically polarized light waves pass through a 45° diagonal polarizer:
   A. Light will pass through with 100% intensity.
   B. Light will pass through with 50% intensity.
   C. Light will not be able to pass through.
   D. This question can only be answered if the light is diagonally polarized.

2. When a stream of diagonally polarized photons passes through a vertical polarizer:
   A. The photons will pass through the polarizer with probability 1.
   B. The photons will pass through the polarizer with probability 0.5.
   C. Each of the photons will break down into two smaller photons. One of the smaller photons will pass through the polarizer and the other will be absorbed.
   D. Only a wave can have polarization, not a particle, hence the photon will not be able to pass through the polarizer.

3. After vertically polarized light of intensity I goes through a 45° diagonal polarizer first and then a horizontal polarizer, the final intensity is:
   A. I.
   B. 4I/3.
   C. I/2.
   D. I/4.

4. If unpolarized light is incident on a double-slit interference experiment with a polarizer after the double-slits, what happens to the interference pattern?
   A. It remains unchanged.
   B. It is halved in intensity.
   C. There is no interference.
   D. Answer depends on the orientation of the polarizer.

5. When a water wave is incident on a double-slit experiment, which slit(s) does the wave pass through?
   A. Neither slit.
   B. Either one of the slits only, creating no interference.
   C. Both slits, thereby creating interference.

6. When particles (bullets) are incident on a double-slit experiment, which slit(s) do the particles pass through?
   A. Neither slit.
   B. Either one of the slits at a time, creating no interference.
   C. Both slits, thereby creating interference.

7. For laser light incident on a single-slit setup with a slit width comparable to the wavelength of the light, you expect to see:
   A. A very intense primary maximum with much smaller secondary maxima to either side.
   B. Equally intense maxima everywhere.
   C. No interference patterns.
   D. No spread of light, only a single narrow line of light the size of the slit.

8. In the double-slit experiment, as the slit separation is increased, the maxima or fringes:
   A. Move closer together.
   B. Maintain the same distance.
   C. Move apart from each other.

### Questionnaire 2: Experiment Part 1 (VV Slits)

*Pass the laser light through the Young's double-slit experiment with the double-slit labelled VV (vertical polarizers before both slits).*

*Predict and Explain Questions (Before carrying out the experiment)*

1. Predict what you expect to see on the screen.

*Observe and Explain Questions (After carrying out the experiment)*

2. Note down what you observe on the screen.
3. Is this observation consistent with your prediction earlier?
4. Explain why you do or do not see the interference pattern.
5. If your prediction and observation do not match, identify why your original prediction was incorrect.

### Questionnaire 3: Experiment Part 2 (VH slits and vertical analyzer)

*Modify the set-up by using the double-slit – VH (containing a vertical polarizer before one of the slit and a horizontal polarizer before the other). Place the analyzer cup in a vertical configuration between the polarized slits and the screen.*

Repeat PEOE questions outlined in Questionnaire 2.

**Questionnaire 4: Experiment Part 3 (VH slits and diagonal analyzer)**
*Continue using the double-slit – VH (containing a vertical polarizer before one of the slit and a horizontal polarizer before the other), but now rotate the analyzer cup to a diagonal position (45° between the polarized slits and the screen).*

Repeat PEOE questions outlined in Questionnaire 2.

**Questionnaire 5: Post-demo questions**
1. When diagonally polarized light waves pass through a horizontal polarizer:
    A. Light will pass through with 50% intensity.
    B. Light will pass through with 20% intensity.
    C. This question can only be answered if the light is diagonally polarized.
    D. Light will not pass through.

2. When a stream of vertically polarized photons is incident on a horizontal polarizer:
    A. The photons will pass through the polarizer with probability 1.
    B. The photons will pass through the polarizer with probability 0.
    C. Each of the photons will break down into two smaller photons. One of the smaller photons will pass through the polarizer and other one will be absorbed.
    D. Only a wave can have polarization, not a particle, hence the photon will be able to pass through the polarizer intact.

3. After vertically polarized light of intensity I goes through a 135° diagonal polarizer and then a vertical polarizer, the final intensity is:
    A. I.
    B. I/4.
    C. I/2.
    D. 2I.

4. An unpolarized light of intensity I is incident on a double-slit interference experiment (with no other polarizers). How would the interference pattern change if the incoming light was instead vertically polarized with intensity I?
    A. It is halved in intensity (I/2) from the original polarized light.
    B. There is no interference.
    C. It remains unchanged.

5. If I can tell which slit an object goes through in a double-slit experiment, it means that:
    A. The object behaves like a wave.
    B. The object behaves like a particle.

6. If you were to send marbles through double-slits (where slit size is sufficient for marbles to pass through), what would you observe at the screen?
    A. No interference.
    B. Interference.
    C. Either A or B depending on whether you know which slit the marbles are going through.
    D. Neither A or B.

7. If you were to send electrons through double-slits (where slit size is sufficient for electrons to pass through, *e.g.* comparable to electrons' de Broglie wavelength), what would you observe at the screen?
    A. No interference.
    B. Interference.
    C. Either option A or B depending on whether the electrons' path through the two-slits is distinguishable or not.
    D. Neither A or B.

## 3. Auxiliary uses for the experimental set-up

Apart from its use in the quantum eraser experiment, the set-up discussed in this work can also be used to demonstrate Malus Law and double-slit interference. Additionally, if there is a need to measure light intensity quantitatively, mobile applications like PhyPhox[R4] can be used when the polarizing cups are placed over the front light sensors of a handphone.